# Twisted bilayer graphene for enantiomeric sensing of chiral molecules

Álvaro Moreno[1], Lorenzo Cavicchi[2], Xia Wang[3], Mayra Peralta[3], Maia Vergniory[3,4], Kenji Watanabe[5], Takashi Taniguchi[6], Pablo Jarillo-Herrero[7], Claudia Felser[3], Marco Polini[8,1], Frank H. L. Koppens [1,9]

[1] ICFO-Institut de Ciencies Fotoniques, The Barcelona Institute of Science and Technology, Castelldefels (Barcelona) 08860, Spain
[2] Scuola Normale Superiore, I-56126 Pisa, Italy
[3] Max Planck Institute for Chemical Physics of Solids, 01187 Dresden, Germany
[4] Donostia International Physics Center, Donostia-San Sebastian, Spain
[5] Research Center for Functional Materials, National Institute for Materials Science, Tsukuba, Japan
[6] International Center for Materials Nanoarchitectonics, National Institute for Materials Science, Tsukuba, Japan
[7] Department of Physics, Massachusetts Institute of Technology, Cambridge, MA 02139, USA
[8] Dipartimento di Fisica dell'Università di Pisa, I-56127 Pisa, Italy
[9] CREA-Institució Catalana de Recerca i Estudis Avançats, 08015 Barcelona, Spain

Abstract

Selective sensing of chiral molecules is a key aspect in fields spanning biology, chemistry, and pharmacology. However, conventional optical methods, such as circular dichroism (CD), encounter limitations owing to weak chiral light-matter interactions. Several strategies have been investigated to enhance CD or circularly polarised luminescence (CPL), including superchiral light, plasmonic nanoresonators and dielectric nanostructures. However, a compromise between spatial uniformity and high sensitivity, without requiring specific molecular functionalization, remains a challenge. In this work, we propose a novel approach using twisted bilayer graphene (TBG), a chiral 2D material with a strong CD peak which energy is tunable through the twist angle. By matching the CD resonance of TBG with the optical transition energy of the molecule, we achieve a decay rate enhancement mediated by resonant energy transfer that depends on the electric-magnetic interaction, that is, on the chirality of both the molecules and TBG. This leads to an enantioselective quenching of the molecule fluorescence, allowing to retrieve the molecule chirality from time-resolved photoluminescence measurements. This method demonstrates high sensitivity down to single layer of molecules, with the potential to achieve the ultimate goal of single-molecule chirality sensing, while preserving the spatial uniformity and integrability of 2D heterostructures.

Introduction

Chirality, the property of a system that cannot be superimposed on its mirror image, is a fundamental concept across biology, chemistry, and physics, manifesting at all natural scales from the electron spin to the structure of galaxies [1]. Chirality is a defining characteristic of most organic molecules that are the basis of life, typically existing in one of two enantiomeric forms. This enantiomeric preference significantly impacts biomedical and pharmacological industries, creating a need for accurate and sensitive methods for chiral resolution [2]. Chirality can only be detected through interaction with another chiral system, leading to polarization-dependent light-matter interactions. Optical techniques such as circular dichroism (CD), circular polarization luminescence (CPL), and Raman optical activity (ROA) are well stablished methods to probe chirality [3], [4], [5]. Such interactions strength is dictated by the dissymmetry factor $g$, which scales with the ratio of molecular size to the wavelength of light: $g \sim a/\lambda \lesssim 10^{-4} - 10^{-2}$. This fundamentally limits the sensitivity of these techniques, particularly for detecting small quantities of molecules.

Significant efforts have been directed towards developing micro- and nano-structures that enhance chiral light-matter interactions [6], [7]. Superchiral light, which enhances chirality beyond that of conventional circularly polarized light [8], [9], has been extensively studied using both chiral [10], [11], [12], [13] and achiral [14], [15] plasmonic metasurfaces, as well as dielectric nanoresonators [16], [17]. Despite enhancement factors as high as $10^{13}$ have been reported [15], these methods often require thick molecular coatings of 10-100s of nanometers due to the non-uniform spatial distribution of the enhancement, highly dependent on the resonator geometry. An alternative strategy involves self-assembling structures composed of chiral biomolecules and metallic nanoparticles (NPs), which exhibit enhanced CD due to the local surface plasmon resonances [18], [19], [20]. The assembly of the chiral nanostructures—and thus the chiroptical signal—is strongly influenced by the presence of the target compound, enabling detection at attomolar sensitivities [20], [21]. These methods only detect the presence of the target molecule though, as the chiroptical signal originates from the assembled structure rather than the molecule itself, and typically require specific target-based designs. Plasmonic coupled CD (PCCD) may offer a promising pathway towards the ultimate goal of single-molecule chirality detection. It has been reported that plasmonic hotspots between NPs can allow the detection of a few ($\lesssim 5$) molecules [22], [23], but the need for precise molecular orientation and positioning control still hinders potential applications. The ideal platform for chirality sensing would not only achieve high sensitivity but also offer tunability and spatial uniformity to facilitate integration, eliminating the need for specific molecular positioning or targeted chemistry.

This work explores a new approach for chirality sensing based on resonant energy transfer (RET), leveraging the fact that many biologically relevant molecules and pharmaceuticals fluoresce in the visible or UV ranges. Energy transfer has been found to be sensitive to the chirality of both donor and acceptor through enantioselective quenching of the emitter in solution of chiral complexes [24], [25], [4], [26], [27]. Instead of a specific quencher molecule, we propose using twisted bilayer graphene (TBG), a chiral 2D material [28], [29], as a tunable enantioselective quencher (Fig. 1-a). We investigate the RET efficiency of chiral molecules to TBG performing time resolved photoluminescence measurements. Spatially resolved measurements provides further statistical significance of the effect, showing that it only depends on the intrinsec chirlality matching of molecule and TBG, but not on the specific spatial location. The deposited layers reach down the single molecule thickness, evidencing the high sensitivity of the proposed method.

### TBG as enantioselective quencher

TBG is inherently chiral when viewed as a three-dimensional system, displaying a strong circular dichroism (CD) peak despite being only two atoms thick [28]. This absorption peak is attributed to the interlayer transition, ocurring at the coupling of the two Dirac cones of the individual layers (Fig. 1-b). The optical response of TBG is governed by both longitudinal ($\sigma_{xx}$) and Hall drag-like ($\sigma_{xy}$) conductivities, shown in Fig. 1-c, the latter reflecting the current induced in one graphene layer due to a perpendicular current in the other layer [29]. The sign of $\sigma_{xy}$ changes upon twist angle reversal, originating the chiral behaviour [30]. Moreover, the transition energy is angle-dependent, enabling the CD tuning over a broad range of energies, including the visible and ultraviolet (UV) spectra [28], where the absorption and emission lines of most biomolecules lay.

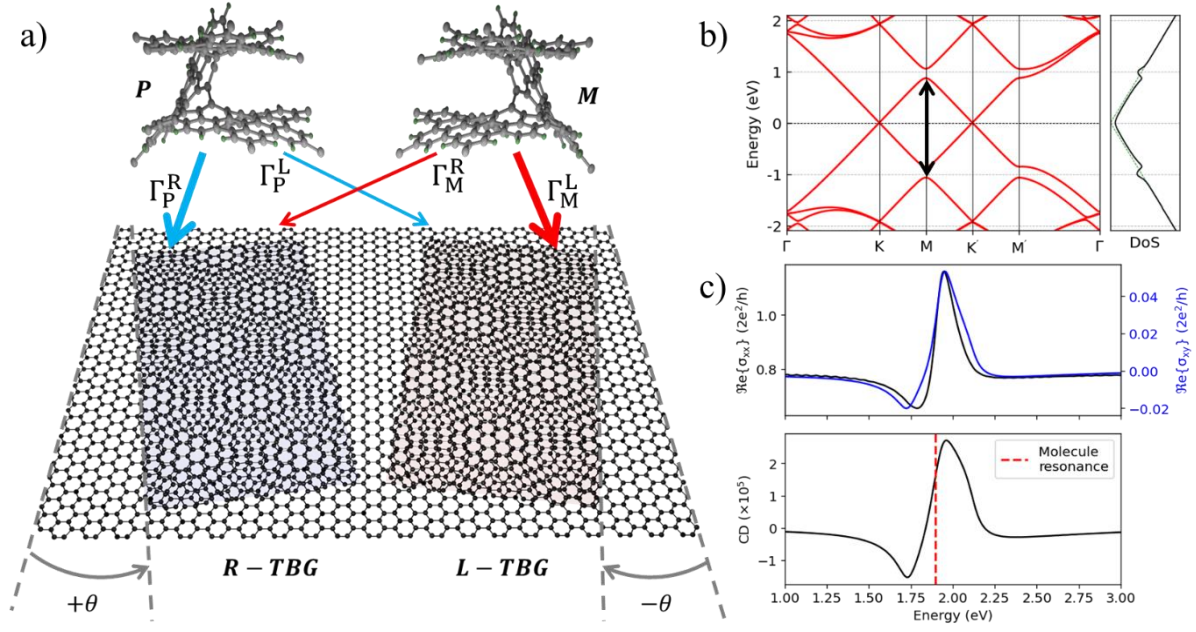

Figure 1: a) Representation of the enantioselective quenching of chiral molecules on TBG. The molecular enantiomers are denoted as P, M, while the TBG chiralities, corresponding to +/- twist angle $\theta$, are labeled R, L for clarity. Energy transfer is more efficient when having matching chiralities (P,R), (M,L), rather than when they are opposite, yielding a higher enhancement in the decay rate of the molecule $\Gamma_{P,M}^{R,L}$. Molecules are represented with a simplified version of the actual structure. b) Band structure of the mini Brillouin zone of hybridized TBG for $\theta = 11°$. The optical resonance is located at the $M, M'$ points, as depicted by the vertical arrow. c) Absorption properties of TBG. The longitudinal ($\sigma_{xx}$) and Hall drag-like ($\sigma_{xy}$) conductivities in the top panel, and the resultant $CD = \Re\{\sigma_{xy}\}/2\Re\{\sigma_{xx}\}$ in the bottom one. The red dashed line marks the molecule transition energy as calculated from TDDFT.

It has already been shown that fluorescent molecules near single layer graphene (SLG) are strongly quenched due to to a non-radiative energy transfer process, resulting in higher decay rates and reduced lifetime [31], [32], [33], [34], [35]. In the dipole approximation, this is retrieved from the power dissipated by an electric dipole $\boldsymbol{\mu}$, $P = P_0 + \frac{\omega}{2}\Im m[\boldsymbol{\mu}^* \cdot \boldsymbol{E}_{sc}(\boldsymbol{r}_0)]$, where $P_0$ is the dissipated power in vacuum and $\boldsymbol{E}_{sc}(\boldsymbol{r}_0)$ is the scattered electric field at the dipole's position. The decay rate enhancement is then given by $\Gamma/\Gamma_0 = P/P_0$, where $\Gamma_0$ is the decay rate in vacuum [36].

In the case of a chiral molecule, the emission characteristics can be modeled as a coupled electric and magnetic dipole moment $\boldsymbol{d}_\pm = \pm i\boldsymbol{\mu} + \boldsymbol{m}$, with a $\pm \pi/2$ phase accounting for helicity [37]. In this case both moments contribute, so the total power dissipated is $P = P_\mu + P_m = \frac{\omega}{2}\Im m[(\pm i\boldsymbol{\mu})^* \cdot \boldsymbol{E}_{sc}(\boldsymbol{r}_0)] + \frac{\omega}{2}\Im m[\boldsymbol{m}^* \cdot \boldsymbol{H}_{sc}(\boldsymbol{r}_0)]$. The fields can be split into contributions from each moment, $\boldsymbol{E}_{sc} = \boldsymbol{E}_\mu + \boldsymbol{E}_m$ and $\boldsymbol{H}_{sc} = \boldsymbol{H}_\mu + \boldsymbol{H}_m$, leading to three components of the dissipated power: $P = P_{\mu\mu} + P_{mm} \pm P_{\mu m}$. The first and second terms correspond to the power dissipated by $\boldsymbol{\mu} \cdot \boldsymbol{E}_\mu$ and $\boldsymbol{m} \cdot \boldsymbol{H}_m$ interactions, while $P_{\mu m}$ contains the crossed terms. For a vertically oriented dipole at a distance $z_0$ above the TBG, $P_{\mu m}$ can be explicitly expressed as:

$$P_{\mu m} = \mp \frac{\omega^2}{2}|\boldsymbol{\mu}||\boldsymbol{m}| \int \frac{d^2\boldsymbol{k}}{(2\pi)^2} \frac{k^2}{k_{z,1}k_1} \Im m\left[e^{2ik_{z,1}z_0} r^{s,p}\right] \qquad (1)$$

Reflecting the molecular ($\boldsymbol{d}_\pm$) chirality in the global sign, and the TBG chirality through the mixed s- and p- polarized Fresnel coefficient $r^{p,s}$, which is linearly dependent on $\sigma_{xy}$. Notably, from the system's symmetry, $P_{\mu m}$ satisfies:

$$P_{\mu m}(\theta, d_+) = -P_{\mu m}(-\theta, d_+) = -P_{\mu m}(\theta, d_-) = P_{\mu m}(-\theta, d_-) \qquad (2)$$

This summarizes the base of the presented approach: energy transfer, and thus the fluorescence quenching, is more efficient when the chirality of molecule and TBG match, and less efficient when they are opposite.

### Chirality signatures on time resolved fluorescence

The device is composed of two TBG areas with the same angle value but opposite signs, resulting in the two chiralities, encapsulated in hBN (illustrated in Fig. 2-a). The use of a thin top hBN layer maximizes the energy transfer efficiency. The emission quenching can reach up to a $10^2$ factor for a spacing $\sim 5\,nm$ [35], necessitating a thickness that allows sufficient radiative emission to enable measurement. We find that a thickness range of 3-5 nm still permits count rates of $10^3$-$10^4$ Hz, which is well above the dark count rate of the detector.

For this work, we use helical bilayer nanographenes (HBNG) as chiral fluorophores, which have attracted considerable attention owing to their chiroptical properties [38]. In particular, [10]HBNG1 is expected to have enhanced configurational stability due to the embedded heptagons [39]. In contrast to previous works, we study the molecules on top of a 2D material instead of in solution. To gain a more accurate understanding of their chiroptical properties, we performed time-dependent density functional theory (TDDFT) calculations of a relaxed molecule on hBN. The calculations predict lowest energy transition at 1.9 eV, in agreement with the experimentally measured emission peak in the molecule emission spectrum. Therefore, we choose the TBG angle to be close to 11° to align the CD peak with the emission energy of the molecules (Fig. 1-c).

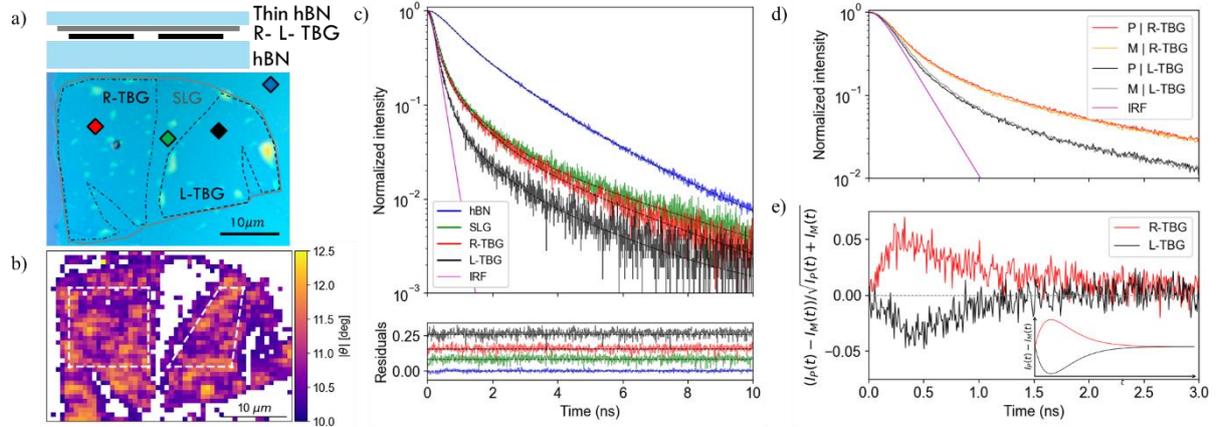

Figure 2: a) Optical image of the sample, where the SLG and the two chiral R and L TBG regions are marked. The diagram above shows the lateral view of the sample, gray and black corresponding to the top and the two bottom rotated SLG respectively, creating the R- and L-TBG. b) Twist angle in the TBG areas retrieved from Raman map. The marked areas are used to extract a mean value and dispersion of the twist angle. c) Normalized fluorescence intensity as a function of time in the different regions of the sample: outside the graphene structure (hBN), on SLG and on both chiralities of TBG. The inset shows the residuals of the fits (dashed lines) normalized to the detector's shot noise. d) Comparison between decay time traces of P and M enantiomers on both R, T-TBG. The difference between the curves is shown in (e).

We deposit a thin layer of homochiral molecules from a solution of [10]HBNG1 in DCM using dynamic spin coating. This layer typically has a thickness of $2.5\,nm$, which is consistent with a 1-2 molecule layer [39]. The top hBN encapsulation allows us to remove the capping layer using a soft oxygen plasma cleaning process [35]. After removing the molecules, we deposit the opposite chirality, so that we can compare the signal from both enantiomers (P, M) on both TBG chiralities (R, L).

In order to estimate the twist angle of the sample we employ resonant Raman spectroscopy. When the excitation energy resonates with an electronic transition, the scattered signal can increase by up to three orders of magnitude [40]. In TBG, the angle-dependent interlayer optical transition leads to a dependency on the Raman G peak enhancement [41], [42]:

$$\frac{I_G^{TBG}}{I_G^{SLG}} \propto \left|\frac{1}{(E_L-E_{Res}(\theta)-i\gamma)(E_L-E_{vHs}(\theta)-\hbar\omega_G-i\gamma)}\right|^2 \quad (3)$$

Here, $E_L = 2.33\ eV$ (532 nm) is the excitation laser energy, $E_{Res}(\theta)$ is the angle-dependent interlayer optical resonance energy, $\hbar\omega_G$ is the energy of the G phonon and $\gamma$ is the resonance window width, associated to the excited-state lifetime. We assume $\gamma = 0.12\ eV$ from other works [41]. This connection enables us to establish a relationship between the TBG angle and $I_G$ peak enhancement, as illustrated in Fig 2-b (Supp. 3). The angle obtained from the two chiral regions (marked areas in Fig 2-b) is very similar, with a value of $11.0 \pm 0.4°$, aligning with the target. At room temperature, the CD peak at this energy has a full-width at half-maximum (FWHM) of 0.2eV (equivalent to 1.2°), thereby offering a suitable avenue for overlapping the chiral effect in TBG with the molecule's resonance.

In order to assess the impact of chirality on FRET, we conducted time-correlated single photon counting (TCSPC) measurements. As depicted in Fig. 2-c, the normalized fluorescence intensity is plotted against time at various locations within the sample, along with the instrument response function (IRF). As anticipated, the molecules exhibit a reduced lifetime when positioned on top of the SLG owing to energy transfer, and they experience further quenching in the TBG region. We observe that the decay curves for R- and L-TBG exhibits significant differences. However, this discrepancy may stem from the spatial heterogeneity within the sample (e.g. differences in stacking quality between R and L sites). To eliminate such errors, we compare the intensities of the two molecular enantiomers (P, M) at the same spatial location (Fig. 2-d). Although the curves are similar, we observe that $I_P^{R-TBG}(t) > I_M^{R-TBG}(t)$, whereas for opposite TBG chirality $I_P^{L-TBG}(t) < I_M^{L-TBG}(t)$, which is more apparent when considering the difference of the curves (Fig. 2-e). This finding is consistent with a chiral effect. Assuming a decay of the form $I_{P,M}^{R,L}(t) = I_0 e^{-t\Gamma_{P,M}^{R,L}}$ with $\Gamma_{P,M}^{R,L} = \Gamma_{achiral} \pm \Gamma_{chiral}$ ($\Gamma_{achiral} > \Gamma_{chiral}$), we qualitatively reproduce the data (Fig. 2-e inset). The sign of the chiral contribution follows equation (2): is positive if we have same molecule-TBG chirality, that is (P, R) or (M, L), and negative otherwise. The measured differences are not completely symmetric with respect 0, though. This may stem from spatial inhomogeneities (twist angle, coating thickness) or differences in the coating between P and M, which may result in changes on the non-radiative rates between the two chosen points. Consequently, we proceed with a quantitative analysis by extracting the lifetime.

### Spatial mapping of chiral effect

To determine the emitter lifetimes, we employed a three-exponential decay model to fit the intensity data: $I_{model}(t) = Ae^{-t/\tau_A} + Be^{-t/\tau_B} + Ce^{-t/\tau_C}$. This model aligns with TDDFT, which predicts the lowest energy transition to have a degeneracy of three. The molecules on top of graphene exhibit strong quenching, resulting in intensity decay curves that are close to the IRF. To accurately determine the lifetimes, we utilized the reconvolution method, which involves fitting the convolution of the IRF and decay model: $I_{conv}(t) = IRF * I_{model}(t)$. This method can retrieve lifetimes down to one order of magnitude smaller than the width of the IRF [43], of $0.4\ ns$ in our case. In Fig. 2-c, we present the fits of the curves along with the residuals, which are the differences between the data and fit normalized by the shot noise, that is $r_i = (I(t_i) - I_{conv}(t_i))/\sqrt{I(t_i)}$, where $t_i$ is the delay of the i-th data point $I(t_i)$. The results for all four points shown in the image and the corresponding ones for the opposite molecule chirality are in Extended Table 1. To improve the accuracy of the fits, we also locally averaged the data within adjacent points for each point.

The focus of our analysis is on the first component of the fit, $\tau_A$, as it holds the most significant weight. Specifically, in areas such as SLG or TBG, the contribution of this factor typically exceeds 95% owing to the strong quenching. Comparing the data from P and M locations, we observe a difference in the values obtained from the hBN and SLG regions. To address this systematic error, we normalize the results using the value at hBN (or SLG), which lacks any chiral effect. Our findings corroborate our expectations, as the results still show the predicted sign change: $\tau_P^R/\tau_P^{hBN} - \tau_M^R/\tau_M^{hBN} = 0.012 \pm 0.002$, while $\tau_P^L/\tau_P^{hBN} - \tau_M^L/\tau_M^{hBN} = -0.032 \pm 0.004$. Analogous results are found when normalizing by SLG, but we choose in the forward to normalize using the hBN values since the signal in the SLG area is more than a 10-fold weaker, and thus noisier. To reinforce our findings, we must demonstrate the spatial homogeneity of the effect, for which we mapped of the emitter lifetimes in the sample for both chiralities.

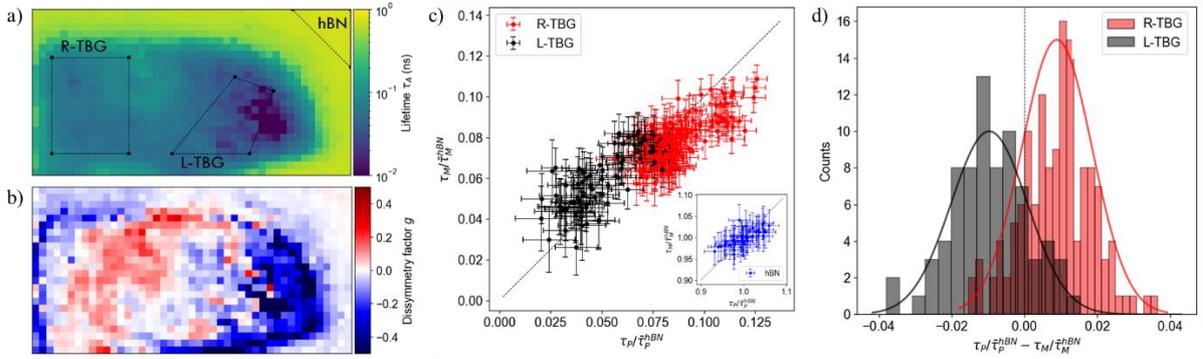

Figure 3: a) Fit lifetime $\tau_A$ map. b) Dissymmetry factor calculated from the maps for the two chiralities. c) Correlation plot of the lifetimes of the two enantiomers P and M for the two chiral TBG and SLG regions, shown in (a). Data is normalized with respect to hBN average value. The inset shows the normalized values for hBN. Error bars are the fit estimated standard deviation. d) Histogram of the differencees between the two enantiomers lifetimes at each chiral region, together with the fit Gaussian envelope. The obtained mean and standard deviation in each case is $\mu_R, \sigma_R = 0.009, 0.010$ and $\mu_L, \sigma_L = -0.010, 0.010$.

Fig. 3-a shows a map of the fit $\tau_A$. By comparing it to the Raman map, we determine the spatial locations of the R- and L-TBG regions. The area outside the sample was used to normalize the lifetime values of the unquenched molecules on hBN. Fig. 2-c shows the correlation plot between the results for both enantiomers. If there were no chiral effect, we may expect a distribution of points centered on the diagonal. However, while the R-TBG region is shifted below the diagonal, the L-TBG is positioned above it. To determine the statistical significance of these results, we calculate the differences of the normalized lifetimes. The resulting histograms are in Fig. 3-d. We find that, in average, in the right-handed TBG $\tau_P^R/\tau_P^{hBN} > \tau_M^R/\tau_M^{hBN}$, whereas in the left-handed TBG $\tau_P^L/\tau_P^{hBN} < \tau_M^L/\tau_M^{hBN}$. Both distributions can be fit to a Gaussian, confirmed by Shapiro-Wilk normality tests, resulting in mean values $\mu_{R,L} = \pm 0.01$ and standard deviations $\sigma_{R,L} = 0.01$. The distribution width is attributed to the angle inhomogeneity. Nevertheless, the number of data points suffices to state that the two distributions have indeed different mean, with a significance level of $p = 0.001$.

The observed differences relate to the dissymmetry factor $g$ of the molecules, a more meaningful factor which characterizes the strength of the chiral interaction [8]. This is usually calculated as the relative intensity difference between the readings of the two enantiomers, which for this case reads:

$$g = \frac{\tau_P/\tau_P^{hBN} - \tau_M/\tau_M^{hBN}}{\tau_P/\tau_P^{hBN} + \tau_M/\tau_M^{hBN}} \quad (4)$$

Fig. 3-b displays a map of the dissymmetry factor values. We observe that the two halves of the sample, corresponding to the R and L TBG areas, have opposite signs, recovering the spatial distribution of TBG chirality. The lack of a clearly resolved SLG area is understood due to its smaller size compared with the other regions, the mapping resolution of 0.75 $\mu m$, and the need for spatial averaging to accurately fit the decay curves, blurring this area. At the edges of the sample, the presence of bubbles and graphite ripples from the stacking gives rise to large inhomogeneities and thus strong differences that are not necessarily related with a chiral effect. Hence only the cleaner inner areas are considered for extracting the histograms. The uniformity and comparability of P and M measurements are evidenced in the area without graphene, where the relative difference is kept below the magnitude of the chiral effect.

### Discussion

The similarity observed between the two distributions in Fig. 3-d highlights the intrinsic symmetry of the system, given the uniformity of the spin-coated sample. The mean value of the differences is equal but opposite in sign, aligning with the theoretical prediction from equation (2):

$$P_{\mu m}(\theta, d_+) - P_{\mu m}(\theta, d_-) = -[P_{\mu m}(-\theta, d_+) - P_{\mu m}(-\theta, d_-)] \quad (5)$$

Additionally, we observe that the dissymmetry factor has values on the order of $g \sim 0.1$. Assuming $\tau_P^{hBN} \approx \tau_M^{hBN}$ for simplicity, the chiral effect model used before $\frac{1}{\tau} = \Gamma_0 \pm \Gamma_{chiral}$ implies that $\Gamma_{chiral} \sim 0.1 \Gamma_{achiral}$, indicating that the chiral contribution constitutes , in average, a 10% of the total decay rate. This result is a hundredfold greater than the theoretically expected values of $g \sim 10^{-3}$, suggesting the presence of an additional mechanism in the near-field interaction between the molecule and TBG, besides photon-mediated energy transfer, that enhances the chiral interaction. Further investigation has to be done in this direction to explain this result.

Overall, our findings position TBG as a promising platform for enantioselective energy transfer, where the quenching of fluorescent molecules is governed by chirality matching. The proposed sensor exhibits a uniform chiral effect by design, with the main parameter determining the strength of the observable effect being the out-of-plane distance between molecule and TBG; this can be controlled with atomic precision by either using 2D materials or atomic layer deposition (ALD) techniques. Such properties make it highly suitable for integration with other technologies. For instance, integrating this method with microfluidic systems could pave the way for real-time, high-throughput analysis of chiral molecules. Moreover, we demonstrated that sensitivities down to the single-molecule layer can be achieved without the need for surface functionalization to target specific molecules. In fact, the approach remains effective even with a minimal number of molecules, with quantum yield being the primary limitation rather than the molecule count. This provides the technique potentiality for single-molecule detection, which would drive insights into fundamental biological processes and advancements in drug development.

# Extended data

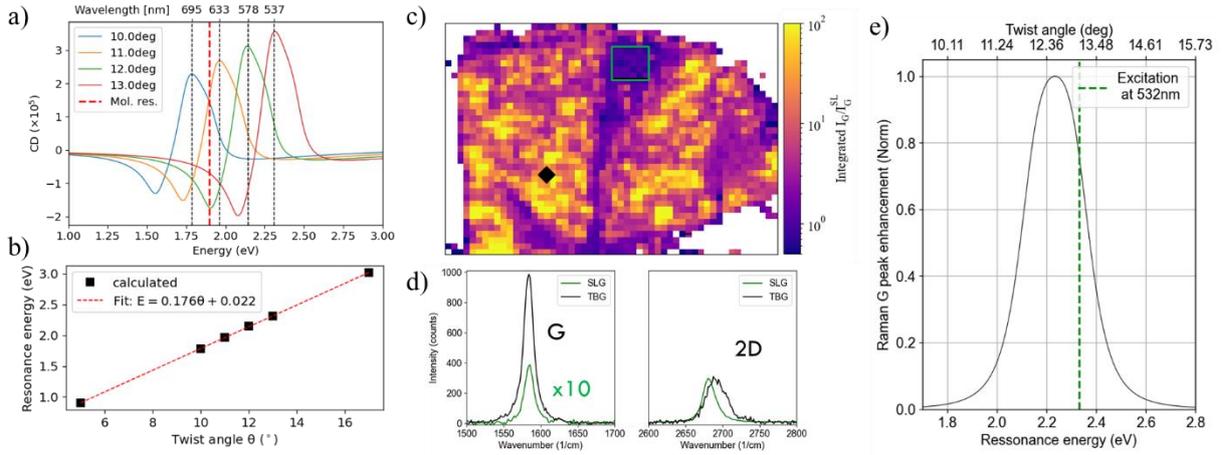

Extended Figure 1: Angle extraction from Raman. a) Calculated CD for different angles in comparison to the calculated lowest energy transition of the molecule. b) Absorption peak energy (transition energy) as a function of the angle. A linear fit is done to extrapolate to any anlgle. c) Map of the ratio between the integrated G peak signal and the mean value of the G peak in the SLG area, marked in the green rectangle. d) Raman G and 2D peaks at the indicated locations. For the SLG, this is the averaged spectrum taken as reference in the map. The SLG G peak has been increased 20 times for clarity. e) Calculated relation between the optical transition energy, or twist angle, and the relative enhancement ratio. The relation between the energy and the angle has been done using the fit in (b). The extraction of the twist angle is done by matching the highest enhancement factor in the map with the maximum of the curve, then assuming that all the angles are in the left side slope of the curve.

| | | A | $\tau_A$ (ns) | B | $\tau_B$ (ns) | C | $\tau_C$ (ns) |
|---|---|---|---|---|---|---|---|
| **P** | hBN | 0.564 ± 0.003 | 0.738 ± 0.004 | 0.407 ± 0.002 | 1.91 ± 0.01 | 0.029 ± 0.001 | 4.28 ± 0.04 |
| | SLG | 0.9684 ± 0.0006 | 0.058 ± 0.001 | 0.0260 ± 0.0006 | 1.028 ± 0.007 | 0.0049 ± 0.0001 | 4.28 ± 0.02 |
| | R-TBG | 0.9629 ± 0.0006 | 0.062 ± 0.001 | 0.0311 ± 0.0006 | 0.938 ± 0.006 | 0.0053 ± 0.0001 | 3.67 ± 0.02 |
| | L-TBG | 0.9933 ± 0.0006 | 0.023 ± 0.003 | 0.0051 ± 0.0006 | 0.894 ± 0.009 | 0.00077 ± 0.00009 | 3.88 ± 0.04 |
| **M** | hBN | 0.543 ± 0.003 | 0.796 ± 0.005 | 0.431 ± 0.003 | 1.98 ± 0.01 | 0.026 ± 0.001 | 4.42 ± 0.05 |
| | SLG | 0.9610 ± 0.0005 | 0.070 ± 0.001 | 0.0335 ± 0.0005 | 0.958 ± 0.005 | 0.00485 ± 0.00008 | 4.28 ± 0.02 |
| | R-TBG | 0.9694 ± 0.0006 | 0.057 ± 0.001 | 0.0259 ± 0.0006 | 0.995 ± 0.006 | 0.0042 ± 0.0001 | 3.88 ± 0.02 |
| | L-TBG | 0.9842 ± 0.0003 | 0.050 ± 0.001 | 0.0136 ± 0.0003 | 0.849 ± 0.007 | 0.00165 ± 0.00004 | 4.07 ± 0.03 |

Extended Table 1: Fit results for the decay curves in the locations of Fig. 2-a. Each decay histogram is normalized and averaged over the adjacent pixels before performing the fit. The fit parameters correspond to the ones of the decay model described in the main text. The uncertainties correspond to the variances of the fit values extracted from the fitting procedure.